\documentstyle[preprint,aps]{revtex}
\begin{document}
\draft

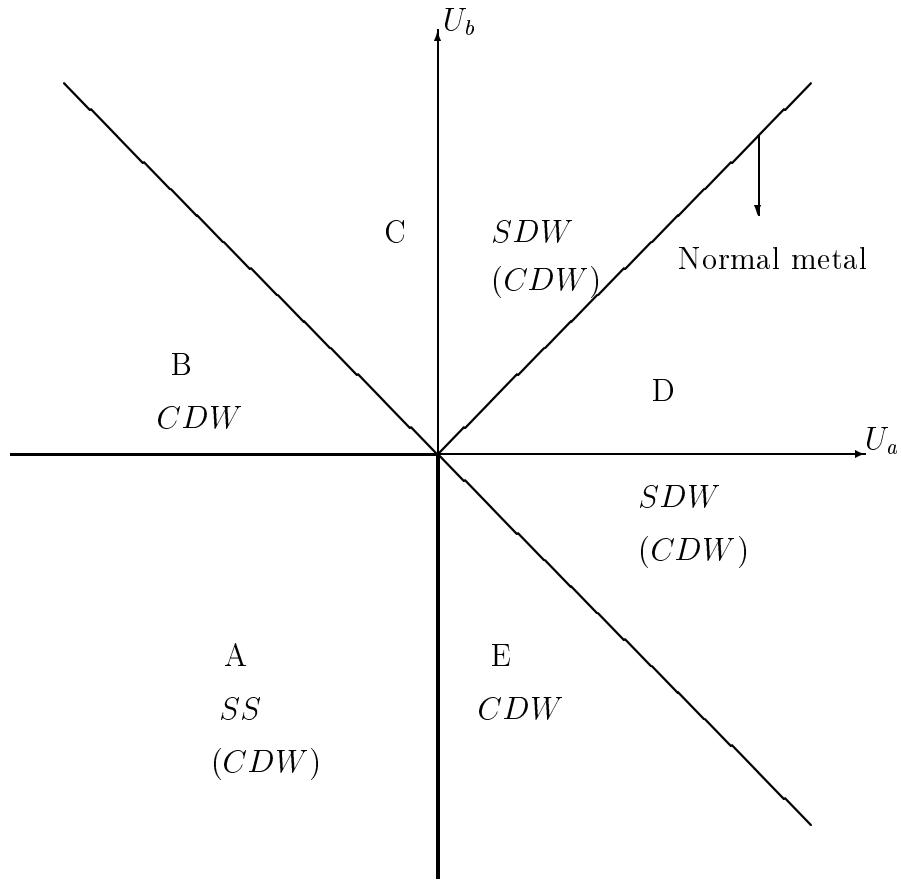
\title{Renormalization group approach to the one-dimensional 1/4-filled
Hubbard model with alternating on-site interactions
}
\author{ G. Jackeli$^{a,b}$, G. Japaridze$^{c}$}
\address{
$^{a}$Joint Institute for Nuclear Research, Dubna, Moscow region,
 141980, Russia \\
$^{b}$ Department of Solid State Physics,
Tbilisi State University, Tbilisi, Georgia\\
$^{b}$ Institute of Physics, Georgian Academy, Tbilisi, Georgia.}

\maketitle
\begin{abstract}\widetext

The one-dimensional Hubbard model with different on-site interactions
is investigated by renormalization group technique.
In the case of a 1/4-filled band the dynamical nonequivalence of sites
leads to the appearance of Umklapp processes in the system
and to the dynamical generation of a gap in the charge excitation
spectrum for $U_{a}\not=U_{b}$, $U_{a}>0$ or $U_{b}>0$.
The ground-state phase diagram is obtained in the
limit of second order renormalization. Depending on the sign and
relative values of the bare coupling constants, there is a gap in the
spin or charge excitation spectrum and the model system tends to
superconducting or antiferromagnetic order at $T=0$, with doubled
period. The role of interaction between particles
on nearest and next-nearest neighbor sites is also considered.
\end{abstract}
\narrowtext

\section{Introduction}

The remarkable discovery of high-Tc superconductivity ~\cite{BM}
in $Cu-O$ compounds has raised the interest in the physics of
correlated electrons in low dimensions. Although the main interest is
centered around the physics of two-dimensional highly correlated
electron systems, the one-dimensional analogues of models
related to high temperature superconductivity are very popular due to
the conjecture ~\cite{P} that properties of the 1D and 2D variants of
certain models have common aspects.

 Emery ~\cite{E1} proposed a one dimensional version  of two-band
copper-oxide model ~\cite{E2}. In Emery's model two essential features
distinguishing the $Cu$ and $O$ sites in real systems are incorporated:
1. the chemical aspect - the hole has different on-site energies on
copper and oxygen sites and 2. the dynamical aspect - there are
different on-site interactions between two holes on copper sites
($U_{a}$) and on oxygen ($U_{b}$). Detailed analysis of the two-band
1/4-filled Peierls-Hubbard model in the large $ U $ limit is given in
~\cite{SBB,PM} and in the case of large dimerization in ~\cite{PM}.
 To investigate the influence of the dynamical nonequivalence of copper
and oxygen sites Japaridze $\it et\/al.$ ~\cite{J} considered a rather
formal 1D version of the $Cu-O$ chain, namely the one-band Hubbard model
with different on-site interaction
on even and odd sites. This model was investigated in the framework
of standard
weak coupling approach using a bosonization technique.

The another method by which insight can be gained
into the problem of interacting
fermions in 1D is the renormalization group approach
~\cite{K,S2,R}. In this paper we consider the quarter-filled
band case of the one-band Hubbard model with alternating
interactions using the renormalization group technique.

As will be shown in \mbox{\bf 2}, the continuum
Hamiltonian of the model has a general form,
describing the one-dimensional system of interacting fermions.
The Hamiltonian of the same form has been considered by
Kimura ~\cite{K} using the conventional field-theoretical
renormalization group. Solyom ~\cite{S2} performed the
second order scaling procedure,
using the bandwidth cutoff, to more general case, when
the forward scattering
process between particles from the same branch was presented.
According to Solyom, taking into account this process substantially
changes results. Unlike ~\cite{S2}, Rezayi $\it et\/ al.$ ~\cite{R}
have shown, that if one modifies the term corresponding
backward scattering and Umklapp process, replacing them by
effective spin and charge carrying interactions, one gets the
theory with momentum transfer cutoff. In this case, the effect
of the forward scattering process between particles
from the same branch is absorbed to a Fermi velocity
renormalization and couplings, which couple the spin-density
degrees of freedom scale independently of the couplings,
which couple the charge-density degrees of freedom.

In our calculation, we shall neglect the
forward scattering process between
particles from the same branch and assuming that
the forces are short range, use the bandwidth cutoff.
In this case, the Lie equations for the coupling constants
decouple into two sets and are analogous to those obtained
by using the momentum transfer cutoff ~\cite{R}.
As will be shown in \mbox{\bf 4 } the scattering
processes responsible to a Fermi velocity renormalization
can only lead to the difference between the degrees of the
divergency of CDW and SDW response functions.

\section{ The model}

The Hamiltonian describing the model has the following form:
\begin{equation}
H=-t\sum_{n,\alpha}c_{n,\alpha}^{+}
(c_{n+1,\alpha}+c_{n-1,\alpha})+
\frac{1}{2}\sum_{n,\alpha}[U+(-1)^{n}V]
\rho_{n,\alpha}\rho_{n,-\alpha}+
\end{equation}
$$
+\sum_{n,\alpha,\beta}[V_{1}\rho_{n,\alpha}
\rho_{n+1,\beta}+V_{2}\rho_{n,\alpha}\rho_{n+2,\beta}]
$$
where $c_{n,\alpha}^{+}(c_{n,\alpha})$ is
a creation (annihilation) operator
for a particle of spin $\alpha$ in the
Wannier state localized at the $n$th
lattice site, $\rho_{n,\alpha}=c_{n,\alpha}^{+}c_{n,\alpha}$ and
\begin{equation}
U=\frac{1}{2}(U_{a}+U_{b})\/, \;\;\;\;\;
V=\frac{1}{2}(U_{a}-U_{b})
\end{equation}
where $U_{a}(U_{b})$ describes on-site interaction on even (odd)
lattice sites and
$V_{1}$ and $V_{2}$ describe the interactions
between particles on nearest and
next-nearest neighbor sites, respectively.

The continuum limit of the Hamiltonian (1) can be taken
by approximating the
spectrum $-2t\cos k$ by linear spectra around $\pm k_{F}$
and using the correspondence:
\begin{equation}
c_{n,\alpha}\rightarrow exp(ik_{F}n)\psi_{1,\alpha}(x)+
exp(-ik_{F}n)\psi_{2,\alpha}(x)
\end{equation}
where the fermion field $ \psi_{1(2),\alpha}(x)$
describes the particles around
Fermi point $ +(-)k_{F}$.

The information about the lattice structure of the
initial model (1) is contained in the oscillating factors
$exp(\pm ik_{F}n)$ and manifests itself in the continuum limit
Hamiltonian via the selection of the scattering
processes relevant for the physics of the system at  large distances.
Except for the cases of quarter- and half-filled bands the dynamical
nonequivalence of sites is irrelevant (in the sense used in connection
with the renormalization group) and the continuum limit Hamiltonian
has the structure of the well known Luther-Emery backscattering model
 ~\cite{LE}, with parameters depending on $U$ ~\cite{J}.
In what follows we will consider the case of quarter-filled band.
In this case the dynamical nonequivalence of sites  becomes
important and  the continuum limit Hamiltonian has the form of
backscattering model supplemented by Umklapp processes with bare
coupling constant proportional to $V$.
\begin{equation}
H=-iv_{F}\sum_{\alpha}^{}\int\limits_{}^{}dx
[\psi_{1,\alpha}^{+}\partial_{x} \psi_{1,\alpha} -
\psi_{2,\alpha}^{+}\partial_{x} \psi_{2,\alpha}]+
\label{u10}
\end{equation}
$$
+\pi v_{F}\sum_{\alpha,\beta}^{}\int\limits_{}^{}dx\Bigl\{
[g_{1\parallel}\delta_{\alpha,\beta}+g_{1\perp}\delta_{\alpha,-\beta}]
\psi_{1,\alpha}^{+}\psi_{2,\beta}^{+}\psi_{1,\beta}\psi_{2,\alpha}+
$$
$$
+[g_{2\parallel}\delta_{\alpha,\beta}+g_{2\perp}\delta_{\alpha,-\beta}]
\psi_{1,\alpha}^{+}\psi_{2,\beta}^{+}\psi_{2,\beta}\psi_{1,\alpha}+
\frac{g_{3}}{2}\delta_{\alpha,-\beta}(
\psi_{1,\alpha}^{+}\psi_{1,\beta}^{+}\psi_{2,\beta}\psi_{2,\alpha}
+\mbox{h.c.})\Bigr\}
$$
The Hamiltonian $(3)$ has a general form describing
the interacting fermion
system in 1D. The terms with coupling constants $g_{1\parallel}$,
$g_{1\perp}$ and $g_{2\parallel}$, $g_{2\perp}$
correspond to the backward and forward scattering processes,respectively
and the terms with $g_{3}$ - to the Umklapp process (see Fig. 1).
In the (3) the processes with coupling constants $g_{1\parallel}$ and
$g_{2\parallel}$ are indistinguishable and
we can choose $g_{2\parallel}=g_{2\perp}$,
without loss of generality. This leads to the following values of the spin
independent bear coupling constants:
$$
 g_{1}=U-2V_{2} \;\;\;\;\;
g_{2}=
U+2V_{1}+2V_{2} \;\;\;\;\;
g_{3}=V
$$
In obtaining (3), the terms corresponding
to forward scattering processes, in which both
incoming particles are from the same branch, were omitted.
For the following discussion the parameter $g_{\rho}=g_{1}-2g_{2}$
is also useful. The coupling constants $g_{s}$ and $(g_{\rho}$,$g_{3}$)
describe the spin and charge degrees of freedom respectively ~\cite{S1}.

\section{ Renormalization group technique }

The multiplicative renormalization group has been used
in the field theory for a long time to eliminate divergences~\cite{BSH}.
This method has two main aspects. The first one is that starting from
a perturbative calculation a partial summation is
obtained by solving the group equations and second
one - a set of equivalent problems can be found which are
described by Hamiltonian of similar form. If there is a model
among the equivalent ones which can be solved, the solution of
the original problem can also be obtained.
The renormalization group treatment of the problem of
interacting fermions in 1D was given by Menyhard and Solyom ~\cite{MS}.
Here we will briefly present the ideas applied to the 1D Fermi gas.

For an unambiguous definition of the model it is necessary to
specify the cutoff procedure which determines the domains of
admissible values of the electron momenta. If the interactions
are short range we have effectively only one limit on momenta and
got a theory with bandwidth cutoff. In the general case, when the
long-range forces are also presented, theory needs two cutoffs,
one as bandwidth and one on momentum transfer~\cite{S1}.
Assuming that the forces are short range in our case,
we shall use the bandwidth cutoff.

At first let us write the auxiliary Green's function $d$
and four-pointed vertex function
$\tilde \Gamma_{i}$ in the following way:
\begin{equation}
G(k_{F},\omega)=d(\frac{\omega}{E_{0}},g_{i})G_{o}(k_{F},\omega)
\end{equation}
\begin{equation}
\Gamma_{\alpha\beta\gamma\delta}(\omega)=
g_{1}\tilde \Gamma_{1}(\omega)\delta_{\alpha\gamma}
\delta_{\beta\delta}-g_{2}\tilde\Gamma_{2}(\omega)
\delta_{\alpha\delta}\delta_{\beta\gamma}
\end{equation}
for process with momentum conversation and
\begin{equation}
\Gamma_{3}=g_{3}\tilde \Gamma_{3}
(\delta_{\alpha\gamma}\delta_{\beta\delta}-
\delta_{\alpha\delta}\delta_{\beta\gamma})
\end{equation}
for Umklapp process,
where $\omega_{0}$ is a bandwidth cutoff.

If the cutoff $E_{0}$ is changed to $E_{0}^ {\prime}$ and
simultaneously the coupling constants $g_{i}$ are
transformed to certain new values
$$
g_{i}^{\prime}=z_{i}^{\prime}g_{i}
$$
then the functions $d$ and $\tilde \Gamma_{i}$ are
multiplicatively transformed to
\begin{equation}
d(\frac{\omega}{E_{0}^{\prime}},g_{i}^{\prime})
=zd(\frac{\omega} {E_{0}^{\prime} },g_{i})
\end{equation}
\begin{equation}
\tilde \Gamma_{i}(\frac{\omega} {E_{0}^{\prime} },
g_{i}^{\prime})=z_{i}^{-1}\tilde
\Gamma_{i}(\frac{\omega} {E_{0}},g_{i})
\end{equation}
Moreover, requiring the invariance of quantity
$g_{i}\tilde \Gamma_{i}d^{2}$, which is an appropriately
renormalized vertex ~\cite{SZ}, only four of renormalization
constants $z_{i}$, $z_{i}^{\prime}$ and $z$ are independent
and we have $z_{i}^{\prime}=z_{i}/z^{2}$.

The scaling equations for $d,\tilde \Gamma_{i}$ and
$g_{i}^{\prime}$ can be written in a common form as:
\begin{equation}
C(\frac{\omega} {E_{0}^{\prime} },g_{i}^{\prime})=
zC(\frac{\omega} {E_{0}},g_{i})
\label{u1}
\end{equation}
For any quantity obeying the condition (\ref{u1})
a Lie differential equation of the form:
\begin{equation}
x\frac{\partial}{\partial x}\ln C(x,g_{i})=
\left.\frac{\partial}{\partial \xi}
\ln C(\xi,g_{i}^{\prime})\right|_{\xi=1}
\end{equation}
can be derived, where $x=\omega/E_{0}$.

The right-hand side of Lie equation can
be calculated by perturbation theory.
If the invariant couplings $g_{i}(x)\equiv g_{i}^{\prime}$
are small for an arbitrary change of the scaling energy
the quantity determined from the solution of the Lie
equation using a few terms of the perturbation series
will present a good approximation in the whole energy range.
If, however, the invariant couplings become
of the order of unity while scaling energy goes
towards lower energies, the right-side of the Lie equation
breaks down and only qualitative results can be obtained.

\subsection{ \mbox{\bf   First-order renormalization} }

Calculating the right-hand side of the Lie equations in the
second-order of the perturbation theory the following
equations for invariant couplings can be obtained
\begin{equation}
x\frac{\partial}{\partial x}g_{s}(x)=g_{s}^{2}(x)
\label{u2}
\end{equation}
\begin{equation}
x\frac{\partial}{\partial x}g_{\rho}(x)=g_{3}^{2}(x)
\label{u3}
\end{equation}
$$
x\frac{\partial}{\partial x}g_{3}(x)=g_{\rho}(x)g_{3}(x)
$$
The first order scaling is equivalent to summing up the
leading terms and these equations are exactly analogous to
those obtained in ~\cite{DL} by parquet approximation.
Let us first analyze the model $(1)$ in the case $V_{1}=V_{2}=0$
In this case the solutions of Eq.(\ref{u2}) and (\ref{u3}) with
boundary conditions $g_{s}(0)=U/\pi v_{F}$,
$g_{\rho}(0)=-U/\pi v_{F}$ and $g_{3}(0)=V/\pi v_{F}$ have the form:
\begin{equation}
g_{s}(x)=\frac{U}{1-Ut}
\label{f1}
\end{equation}
where $t=\ln x$,
\begin{equation}
g_{\rho}(x)=\frac{-U}{1+Ut} \;\;\;\;\;\;
g_{3}(x)=\frac{V}{1+Ut}
\label{f2}
\end{equation}
for $U^{2}=V^{2}$, while for $U^{2}>V^{2}$ we have
\begin{equation}
g_{\rho}(x)=-D\coth D(t_{0}-t) \;\;\;\;\;\;
g_{3}(x)=\frac{(\mbox{sgn} V)D}{\left|\sinh D(t_{0}-t)\right|}
\label{f3}
\end{equation}
where $t_{0}=1/ D\mbox{arccoth} (U/D)$ and for $U^{2}<V^{2}$ they reads as
\begin{equation}
g_{\rho}(x)=-D\cot D(t_{0}^{\prime}-t) \;\;\;\;\;
g_{3}(x)=\frac{(\mbox{sgn} V)D}{\left|\sin D(t_{0}-t)\right|}
\label{f4}
\end{equation}
where $t_{0}^{\prime}=1/D\mbox{arccot} (U/D)$.

These equations show that, at certain values of bare couplings, the poles
appear in the expressions of the invariant coupling constants.
This singular behavior of invariant couplings doesn't indicate the
phase transition at finite temperature, which is impossible in one
dimensional system ~\cite{L}. The existence of the poles in expressions
(\ref{f1})-(\ref{f4}) indicates that at low temperatures the interactions
become strong and this approximation is no longer valid.
In ~\cite{DL} Dzyaloshinsky and Larkin, comparing their result to exact
solution obtained by Gauden ~\cite{G}, emphasized, that the poles in the
expressions of the invariant couplings can indicate  the appearance of
the gap in the one particle excitation spectrum.
Thus we can expect that the spin (charge)  gap exists when there is a
pole in the expression for $g_{s}$ $((g_{3},\mbox{ }g_{\rho}))$.
Thus the charge gap exists in all sectors except \mbox{\bf A},
while the spin gap exists in the sectors
\mbox{\bf A}, \mbox{\bf B}, \mbox{\bf E} (see Fig. 2).

\subsection{ \mbox{\bf  Second-order renormalization} }

The second-order renormalization corresponds to considering next
to leading logarithmic corrections and thereby to taking into
account fluctuation effects. In the limit of third order of
perturbation theory for $d$ and $\tilde \Gamma_{i}$, the following
Lie equations for invariant couplings can be obtained:
\begin{equation}
x\frac{\partial}{\partial x}g_{s}(x)=g_{s}^{2}(x)
[1+\frac{1}{2}g_{s}(x)]
\label{L1}
\end{equation}
\begin{equation}
x\frac{\partial}{\partial x}g_{\rho}(x)=g_{3}^{2}(x)
[1+\frac{1}{2}g_{\rho}(x)]
\label{L2}
\end{equation}
\begin{equation}
x\frac{\partial}{\partial x}g_{3}(x)=g_{\rho}(x)g_{3}(x)[1+
\frac{1}{4}g_{\rho}(x)]+\frac{1}{4}g_{\rho}^{3}(x)
\label{L3}
\end{equation}
The solutions of these equations can be obtained only in an implicit form.
They are the smooth functions of energy and tend to the saturation values,
fixed points, at $\omega=0$. The values of the fixed points can
be obtained from zeros of the right-hand sides of the Lie equations
(\ref{L1})-(\ref{L3})
using the following arguments:
$$
a).\;\;     x\frac{\partial}
{\partial x}g_{s}(x)\leq0\mbox{  for }g_{s}(x)\geq-2;
$$
$$
b).\;\;
x\frac{\partial}{\partial x}g_{\rho}(x)\leq0
\mbox{  for }g_{\rho}(x)\geq-2
\mbox{  and  }\frac{g_{\rho}^{2}(x)-g_{3}^{2}(x)}{g_{\rho}(x)+2}=c,
\mbox{  where }c\mbox{ is a constant.}
$$
The values of the  fixed points depend on the relation
between the bare coupling constants and
 are presented in \mbox{\bf Table 1.}

\subsection{ \mbox{\bf  Response functions and ground-state phase diagram} }

The symmetry of the ground-state can be found by investigation of various
response functions which are expected to be singular.
These quantities are the charge-density (CDW) and spin-density
(SDW) waves with wave-vector $k=2k_{F}$,
the singlet-superconductor (SS) and triplet-superconductor
(TS) response functions
with wave-vector $k=0$ ~\cite{SR}.

In the case of $1/4$-filled band the periodicity of
the CDW and SDW is equal four lattice
constant and it is possible to introduce two separate
sets of order parameters describing the fluctuations of
the charge and spin density on even ($a$) and odd ($b$)
sublatticies. In the real space and time representation
general formula for these functions is
\begin{equation}
\Phi(x,t)=-i\theta(t)<[U(x,t)U^{+}(0,t)]>
\end{equation}
where $ U(x,t)=\sum_{\alpha}^{}[\psi_{1,\alpha}^{+}(x,t)
\psi_{2,\alpha}(x,t)+\psi_{2,\alpha}^{+}(x,t)\psi_{1,\alpha}(x,t)]$
for the charge-density response function $N(x,t)$;
$\psi_{1,\uparrow}^{+}(x,t)\psi_{2,\downarrow}(x,t)+
\psi_{2,\uparrow}^{+}(x,t)\psi_{1,\downarrow}(x,t)$
for the  spin-density response function  $\chi(x,t)$ and
$\psi_{1,\uparrow}^{+}(x,t)\psi_{2,\downarrow(\uparrow)}(x,t)$
for singlet (triplet)-superconductor response
function $\Delta_{s(t)}(x,t)$

The Fourier-transformed response functions don't
satisfy the criterion of multiplicative renormalization,
Solyom ~\cite{SR} introduced the auxiliary renormalizable
functions defined by
$$
\overline{\Phi}_{i}(\omega)=\frac{\partial\Phi_{i}
(\omega)}{\partial\ln (\omega)}
$$
where $\Phi_{i}(\omega)$ is one of the above considered
response functions. In the limit of second order of
perturbation theory
the following Lie equations for these renormalizable response
functions can be obtained:
\begin{equation}
\frac{\partial\overline{N}_{a(b)}(x)}{\partial\ln (x)}=
2g_{s}(x)-g_{2}(x)+(-)g_{3}(x)+F(x)
\label{R1}
\end{equation}
\begin{equation}
\frac{\partial\overline{\chi}_{a(b)}(x)}{\partial\ln (x)}=
-g_{2}(x)-(+)g_{3}(x)+F(x)
\label{R2}
\end{equation}
\begin{equation}
\frac{\partial\overline{\Delta}_{s}(x)}{\partial\ln (x)}=
g_{s}(x)+g_{2}(x)+F(x)
\label{R3}
\end{equation}
\begin{equation}
\frac{\partial\overline{\Delta}_{t}(x)}{\partial\ln (x)}=
g_{2}(x)-g_{s}(x)+F(x)
\label{R4}
\end{equation}
where
$$
F(x)=\frac{1}{2}(g_{s}^{2}(x)+g_{2}^{2}(x)-
g_{s}(x)g_{2}(x)+\frac{1}{2}g_{3}^{2}(x))
$$
and $\overline{N}_{a}(x)$ and $\overline{N}_{b}(x)$
($\overline{\chi}_{a}(x)$ and $\overline{\chi}_{b}(x)$)
describe the CDW (SDW) located on $a$
and $b$ sublatticies , respectively and they are
different only in a sign of $g_{3}$,
the changing $a\ss b$ is equivalent to change
the sign of Umklapp processes.

In the limit $\omega\ss0$ the asymptotic expressions
of response functions are obtained
as
$$
\Phi_{i}(\omega)\propto\overline{\Phi}_{i}
(\omega)\propto(\frac{\omega}{E_{0}})^{\alpha_{i}}
$$
The critical exponents
$\alpha_{i}$ can be obtained by putting the fixed
points of invariant couplings in the
right-hand side of Eqs.(\ref{R1})-(\ref{R4}) and they are presented
in \mbox{\bf Table 2}

According to singularities in the response functions
let us summarize the ground-state phase diagram.
In the sector $\mbox{\bf A}$ the CDW and SS response
functions are divergent.
The dominant singularity is in the
singlet-superconductor response.
In the sectors $\mbox{\bf B}$ and $\mbox{\bf E}$
only CDW response, located on the sublattice with lower
on-site interaction, is divergent. There is coexistence
of CDW and SDW in the sectors  $\mbox{\bf C}$ and $\mbox{\bf D}$.
The CDW is still located on the sublattice with lower on-site
interaction while the SDW is located on the other one.

In the case of nonalternating chain, $g_{3}=0$, $g_{4}$
does not renormalize and system has
the properties of normal metal for $U>0$ and
singlet-superconductor for $U<0$.

We note ground-state phase diagram obtained here is exactly
analogous to that obtained by boson representation theory in ~\cite{J}.

\section{  Scaling to the exactly soluble models}

The renormalization group and scaling arguments presented
in \mbox{\bf 3} establish a relationship between the original problem
and a set of problems in which the coupling constants have the
somewhat different values. If an exactly soluble
model appears among these equivalent systems,
the physical behaviors of the original model can be
predicted by using the scaling arguments.

From Tab.1 one can easily see that depending on relation
between the bare couplings the model can be scaled to
Tomonaga ~\cite{T} or Lutter-Emery (LE) ~\cite{LE} model and the
ground-state phase diagram can be obtained by
using their results ~\cite{S1}.

\mbox{\bf (i).} In the sector $\mbox{\bf A}$  (see Fig. 2),
$g_{s}(1)<0,\mbox{  }g_{\rho}(1)\geq|g_{3}(1)|$, the spin part
of the Hamiltonian scales to the LE line and
there is a gap in the spin part of the excitation spectrum.
The charge part scales to a Tomonaga model and the Umklapp
processes have no influence. Only the CDW and SS responses
can be divergent. The dominant singularity is in the
latter.

\mbox{\bf (ii).} For $g_{s}(1)>0,\mbox{  }g_{\rho}(1)<|g_{3}(1)|$,
sectors $\mbox{\bf C}$, $\mbox{\bf D}$, the situation is reversed
concerning the spin and charge parts of the Hamiltonian.
The gap is in the charge part and CDW and SDW
responses show a singular behavior. The latter is more
singular. The difference between singularity in CDW and
SDW responses is caused
by Fermi velocity renormalization ~\cite{S1} for which
scattering processes omitted in (2) is responsible ~\cite{R}.

\mbox{\bf (iii).} In the sectors $\mbox{\bf B}$ and
$\mbox{\bf E}$, $g_{s}(1)<0,\;\;g_{\rho}(1)<|g_{3}(1)|$,
the Hamiltonian scales to the LE line, there
is a gap in both charge and spin parts and only
charge-density response function
is divergent.

\subsection{  \mbox{\bf The role of intersite interactions}}

Taking into account  the effects of intersite interactions
leads to the renormalization of bare coupling constants Eq.(4).
Let us consider the phase diagram for the arbitrary sign of
the bare coupling constants, not restricting ourselves
by realistic case, when $U_{a}$, $U_{b}$, $V_{1}\mbox{ and },V_{2}>0$.

In the case of extended model the  spin and
charge gaps exist in the case when
$U-2V_{2}<0$ and $-(U+4V_{1}+6V_{2})\geq |V|$, respectively.

\mbox{\bf (i).} Now the situation discussed in
$\mbox{\bf 4.(i)}$ takes a place
when
$$
U-2V_{2}<0 \mbox{  and  } -(U+4V_{1}+6V_{2})\geq  \left|V\right|
$$
and the ground-state of the system is the same as in sector \mbox{\bf A}.

\mbox{\bf (ii).} When the bare couplings constants satisfy the condition:
$$
U-2V_{2}\geq0\mbox{  and  }-(U+4V_{1}+6V_{2})<\left|V\right|.
$$
in the extended model, the same response functions as
in sectors \mbox{\bf B} and \mbox{\bf E} are divergent.

\mbox{\bf (iii).} The system has  the
same ground-state as in sectors
\mbox{\bf C} and \mbox{\bf D} when
$$
U-2V_{2}<0\mbox{  and  }-(U+4V_{1}+6V_{2})<\left|V\right|.
$$

\mbox{\bf (iiii).} Unlike to the case $V_{1}=V_{2}=0$,
when intersite interactions are presented,
the Hamiltonian scales to the Tomonaga model when:
$$
U-2V_{2}\geq0 \;\; \mbox{  and  }\;\; -(U+4V_{1}+
6V_{2})\geq\left|V\right|
$$

There is no gap in either the charge or
spin parts of the Hamiltonian and
the singlet- and triplet-superconductor responses are divergent.

\section{Summary}

The one-dimensional 1/4-filled Hubbard model with alternating on-site
interactions has been considered in the weak coupling
approach using the
renormalization group technique. The model is characterized
by the decoupling of charge and spin degrees of freedom,
but the dynamical nonequivalence of sites leads to the
appearance of Umklapp processes in the system and
to the dynamical generation of a gap in the charge excitation spectrum
for $U_{a}\not=U_{b}$, $U_{a}>0$ or $U_{b}>0$.
The ground-state phase diagram is obtained in the
limit of second order renormalization.
Depending on the sign and relative values of the bare
coupling constants, there is a gap in the spin or
charge excitation spectrum and the model system tends
to superconducting or antiferromagnetic
order at $T=0$, with doubled period.
It is shown by scaling to exactly soluble
models that the terms corresponding to scattering processes,
in which both
incoming particles are from the same branch
lead to the difference
between the degrees of the divergency of CDW and
SDW response functions.
The role of interaction between particles
on nearest and next-nearest neighbor sites is also considered.

\acknowledgments

This work was supported by INTAS grant N 94-3862.

\begin{table}\widetext
\caption{ The values of the fixed points
for invariant coupling constants.
The letters A...C describe the same regions as in Fig. 2 and
  $\gamma=(c+\surd\overline{c^{2}+8c})/2$}
\begin{tabular}{lccccc}
 &A  &B&C&D&E \\ \tableline

$g_{s}(0)$&-2           &-2  &0 &0 &-2 \\ 
$g_{2}(0)$&$-1-\gamma/2$& 0  &1 &1 &0   \\ 
$g_{3}(0)$& 0           & 2  &2 &-2&-2 \\ 
$g_{\rho}(0)$&$\gamma$     &-2  &-2&-2&-2  
\end{tabular}
\end{table}

\begin{table}\widetext
\caption{The critical exponents $\alpha_{i}$}
\begin{tabular}{lccccc}
 &A  &B&C&D&E \\ \tableline
$N_{a(b)}$   &-3/2+$\gamma/2$ &1 (-3)&5 (-3/2)&-3/2 (5/2)&-3 (1)\\ 
$\chi_{a(b)}$&5/2+$\gamma/2$  &1 (5)&-3/2 (5/2)&5/2 (-3/2)&5 (1)\\ 
$\Delta_{s}$ &-3/2-$\gamma/2$ &1&5/2&5/2&1\\ 
$\Delta_{t}$ &5/2-$\gamma/2$  &1&5/2&5/2&1\\ 
\end{tabular}
\end{table}
\begin{figure}
\caption{Classification of the scattering
processes when interaction
terms are expressed in a basis of right (solid lines)
and left (dashed lines).
\rm{\bf a)} Forward scattering in the vicinity of
different Fermi points,
\rm{\bf b)} Backward scattering,
\rm{\bf c)} Umklapp processes.}
\end{figure}

\begin{figure}
\caption{The ground-state phase diagram of
the model (1) $(V_{1}=V_{2}=0)$.
The response functions corresponding to the phases
shown in the parentheses
have a lower degree of divergence than the others.}
\end{figure}

\end{document}